# Numerical simulation studies of nano-scale Surface plasmon components: waveguides, splitters and filters

Xian-Shi Lin and Xu-Guang Huang*
Laboratory of Photonic Information Technology, South China Normal University
Guangzhou, 510006, China

## ABSTRACT

In this paper, we theoretically and numerically demonstrate a two-dimensional Metal-Dielectric-Metal (MDM) waveguide based on finite-difference time-domain simulation of the propagation characteristics of surface plasmon polaritons (SPPs). For practical applications, we propose a plasmonic Y-branch waveguide based on MDM structure for high integration. The simulation results show that the Y-branch waveguide proposed here makes optical splitter with large branching angle (~180 degree) come true. We also introduce a finite array of periodic tooth structure on one surface of the MDM waveguide which is in a similar way as FBGs or Bragg reflectors, potentially as filters for WDM applications. Our results show that the novel structure not only can realize filtering function of wavelength with a high transmittance over 92%, but also with an ultra-compact size in the length of a few hundred nanometers, in comparison with other grating-like SPPs filters. The MDM waveguide splitters and filters could be utilized to achieve ultra-compact photonic filtering devices for high integration in SPPs-based flat metallic surfaces.

## 1. INTRODUCTION

Surface plasmons are waves that propagate along a metal-dielectric interface with an exponentially decaying field in the both sides [1,2]. Waveguides based on surface plasmons, using materials with a negative permittivity such as the noble metals, have the ability to confine and propagate electromagnetic energy in a subwavelength while overcoming the diffraction limit in conventional optics, thus opening up a wealth of new possibilities as sub-wavelength optical components based on metallic photonic materials. For example, thin metal films of finite width embedded in a dielectric which is known as Metal-Dielectric-Metal (MDM) sandwich structure can be used as waveguides to guide the plasmonic signals [3-5]. To date, several novel plasmonic waveguide based MDM structure have been theoretically proposed or experimentally demonstrated, such as U-shaped waveguides [6], T-shaped waveguides as splitters [7,8], Y-shaped combiners and splitters [9,10], modulators and switches [11], couplers [12], reflectors [13], M-Z interferometers [14], Bragg mirrors [15], and photonic bandgap structures [16]. And different techniques have been applied to the fabrication of such waveguides including standard UV lithography and direct laser writing.

In Section 2 of this paper, we present the theory and simulation for the fundamental mode propagating along a two-dimensional MDM slit waveguide based on metal waveguide theory combining with Drude mode.

In Section 3, we investigate a wide-angle plasmonic Y-branch waveguide as power splitter based on direct bending structures, using the method of Finite-Difference Time-Domain (FDTD) and perfect-matching-layer absorbing boundary conditions. The SPPs distributions and propagations of the Y-branch with different branching angles are characterized in detail. Defects of the joint as well as the fabrication tolerance are all addressedd.

In Section 4, we introduce a finite array of periodic teeth structures on one surface of the MDM waveguide structure. A model based on the scattering matrix theory will be given. The spectrum of transmittance of the teeth-shaped structures is simulated with FDTD method. The dependences of the forbidden bandwidth on the period and the period number of the teeth structures will be discussed.

The paper ends with conclusions in Section 5.

*Corresponding author: huangxg@scnu.edu.cn; phone 86-20-39310015;

## 2. SURFACE PLASMON WAVEGUIDES THEORY AND SIMULATION

In this section we will provide basic theory and simulation which are necessary for the understanding of surface plasmon waveguides. For two close metal plates, the fundamental TM mode in the slit transforms to SPPs modes on the metallic surfaces. The dispersion equation for fundamental TM mode in the waveguide is given by [5]

$$\varepsilon_d k_{z2} + \varepsilon_m k_{z1} \coth(-\frac{ik_{z1}}{2}w) = 0, \tag{1}$$

with $k_{z1}$ and $k_{z2}$ defined by momentum conservations:

$$k_{z1}^2 = \varepsilon_d k_0^2 - \beta^2, \quad k_{z2}^2 = \varepsilon_m k_0^2 - \beta^2. \tag{2}$$

Where $\varepsilon_d$ and $\varepsilon_m$ are respectively dielectric constants of the dielectric media and the metal, $k_0=2\pi/\lambda_0$ is the free-space wave vector. The propagation constant $\beta$ is represented as effective index $n_{eff}=\beta/k_0$ of the waveguide for SPPs. The dependence of $n_{eff}$ on the wavelength of the beam and on the width of the slit is shown in Fig. 1. It can be seen that the real part of $n_{eff}$ decreases rapidly with the increase of wavelength from 400nm to 800nm, and then becomes saturation. Meanwhile, it also decreases with increasing the width in Fig. 1(a). The decreasing rate becomes smaller and smaller when the width of the slit is over 20nm. It means the beam will be more favorably to be confined in a narrow area than in a wide area. The imaginary part of $n_{eff}$ is referred to the propagation length which is defined as the length over which the power carried by the wave decays to 1/e of its initial value: $L_{spps}=\lambda_0/[4\pi\cdot\text{Im}(n_{eff})]$. The imaginary part of $n_{eff}$ increases with wavelength after 700nm, and decreases with the increase of the width of a slit waveguide. It is clearly that smaller slit will have larger imaginary part of $n_{eff}$, which leads to higher loss and shorter propagation length as shown in Fig. 1(b), although the field confinement of the slit waveguide is stronger.

In the calculation above and the following simulations, the frequency-dependent complex relative permittivity of silver is characterized by Drude model:

$$\varepsilon_m(\omega) = \varepsilon_\infty - \frac{\omega_p^2}{\omega(\omega+i\gamma)}. \tag{3}$$

Here $\omega_p = 1.38\times10^{16}\,Hz$ is the bulk plasma frequency, which represents the natural frequency of the oscillations of free conduction electrons. $\gamma = 2.73\times10^{13}\,Hz$ is the damping frequency of the oscillations, $\omega$ is the angular frequency of the incident electromagnetic radiation, $\varepsilon_\infty$ stands for the dielectric constant at infinite angular frequency with the value of 3.7 [17].

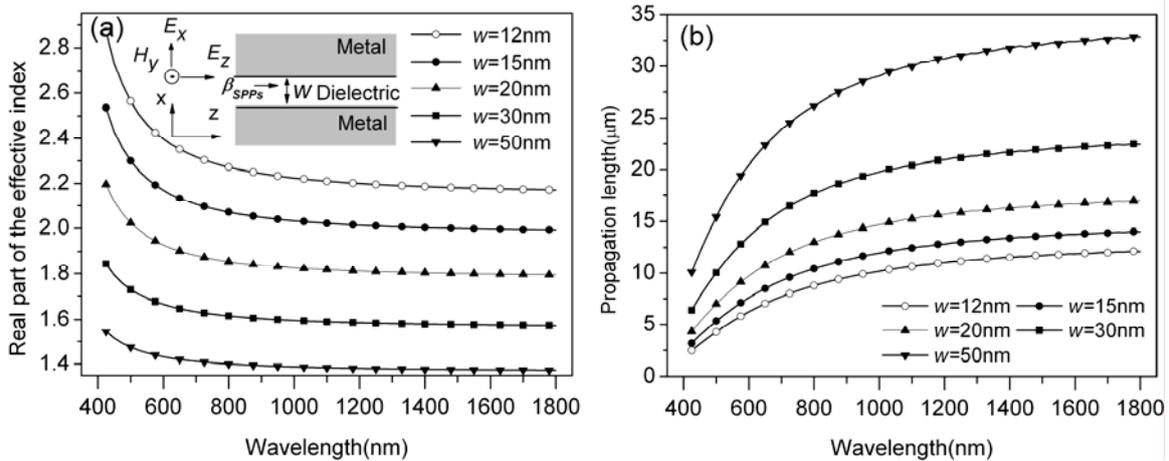

Fig. 1. (a) Real part of the effective index of refraction versus the width of a MDM slit waveguide structure. (b) Propagation length as a function of wavelength with different width of a MDM slit waveguide structure.

# 3. PLASMONIC WAVEGUIDE SPLITTERS

Y-branch waveguides are widely used in power splitter, switches, and combiners etc. However, the conventional dielectric Y-branch waveguides that has a branching angle small than 10°, and with a large size because the separation of the interacting waveguide arms demand very long distances, but also sensitive to the fabrication errors. These characteristics result in the difficulty of fabrication and high integration. In order to achieve nano-scale integration, one way to realize splitting function of the guided plasmons is to use a T-splitter or a wide-angle Y-branch with subwavelength arm-length. A Mach-Zehnder interferometer composed of a Y-branch followed by two parallel waveguides sections has been fabricated and characterized [18]. The application of Y-splitters to branch the SPPs guided by dielectric-loaded SPP waveguides has recently been analyzed theoretically [19]. Recently, a real Y-splitter based on S-shaped bending taking into account fabrication defects, MMI-based and coupler-based splitters with S-shaped bending have been presented [20]. However, most of the splitters mentioned above were based on S-shaped bending structures, while the direct bending structures were not considered so far.

Here we propose a wide-angle MDM plasmonic Y-branch waveguide based on direct bending structure with subwavelength arm-length, as shown in Fig. 2(a). The width of the waveguide is $w$=50nm, the length of $L$ is fixed to be 250nm, and $\theta$ stands for the branching angle. For communication application, we set the input wavelength $\lambda$=1550nm, and the corresponding complex relative permittivity of silver $\varepsilon_m = 0.514 + 10.8i$ [21]. In the FDTD simulation, the grid sizes in the $x$ and $z$ directions are chosen to be 5nm×5nm. The fundamental TM mode of the plasmonic waveguide is excited by a dipole source. The incident power of $P_{in}$ and the transmitted power of $P_{out}$ which are calculated by integrating the normal component of the time-averaged Poynting vector of $\vec{S}_{avg} = \frac{1}{2}\text{Re}(\vec{E} \times \vec{H}^*)$. Here the output fields are well separated and there is no overlap between the modes of the two branches and $P_{out1}$= $P_{out2}$= $P_{out}$ for this case. The transmittance is defined to be $T$= $P_{out}/P_{in}$.

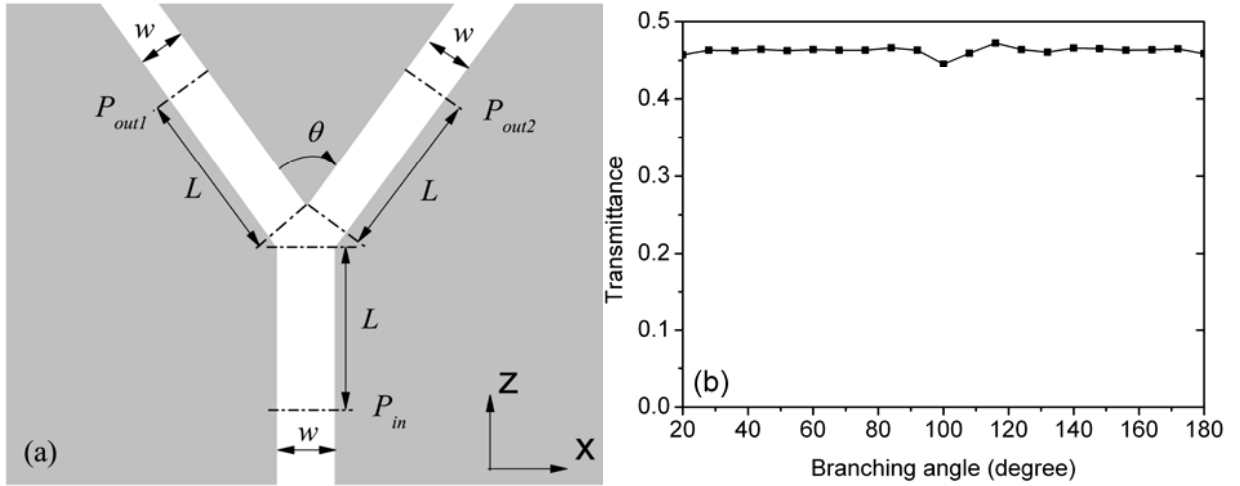

Fig. 2. (a) Schematic of the Y-branch structure studied here. (b) The transmittance of the Y-branch structure under different splitter angles at the wavelength of 1550nm.

Fig. 2(b) shows the transmittance of the Y-branch waveguide under different branching angles at the wavelength of 1550nm. As seen the transmittance keep with a high value over 45% with a wide range branching angle even as large as 180° that becomes a T-shaped waveguide, which is distinguished from the Y-splitter based on dielectric waveguides. The results reveal that the transmittance is not sensitive to the branching angle, and then we can realize optical splitter with random branching angle.

Considering fabrication errors, a non-perfect Y-branch waveguide with a defect at the joint of the branching is introduced as show in Fig. 3(a). Here all the waveguides have the same parameters as the previous one. Figure 3(b) shows the transmittance of the non-perfect Y-branch waveguide for a given branching angle $\theta$=60°. It reveals that the defect in the Y-joint has almost no effect on the transmittance. The fabrication tolerance of the wide-angle plasmonic Y-branch waveguide is quite large that can reduce the requirement of the manufacture equipment.

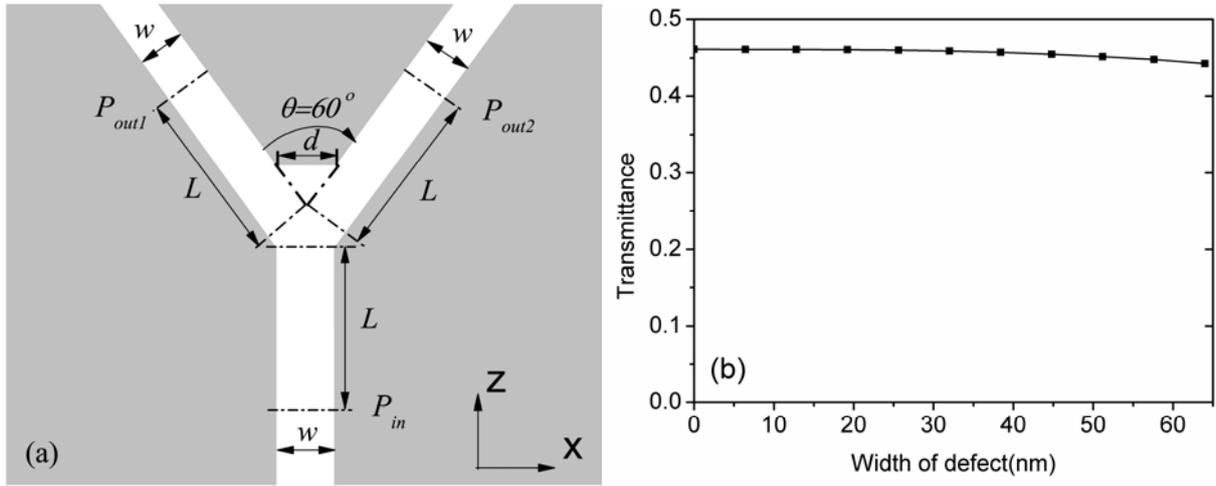

Fig. 3. (a) Schematic of the Y-branch structure with defect. (b) The transmittance of the structure under different defect width.

## 4. AN ULTRA-COMPACT PLASMONIC WAVELENGTH FILTER

As one of the basic components to high integration in photonic circuits, a wavelength filter has to be developed. To achieve the filtering function, several SPPs Bragg reflectors and nanocavities have been proposed. They include a planar heterowaveguide constructed by alternately stacking two kinds of MDM waveguides with the refractive index periodically modulated [22,23], the quasi-periodic case and a double-band plasmon Bragg reflector [24], a low-loss plasmonic Bragg reflector which is constructed by periodic changes in the dielectric materials of the MDM waveguides [25], a structure with periodic variation of the width of the dielectric in MDM waveguides [17]. Most recently, a modeling and design methodology based on characteristic impedance for plasmonic waveguide with MDM structures [26], a high-order plasmonic Bragg reflector with a periodic modulation of the core index of the dielectrics [27] and a wide bandgap plasmonic Bragg reflector [28] are proposed. Most of the structures mentioned above, one of the shortcomings is that the period number of N>9 with the total lengths over 4μm beyond subwavelength-scale, and it is hard to fabrication. To deal with these problems, we propose a wide bandgap plasmonic filter consisting of finite array of periodic teeth structures in nano-scale based on MDM waveguide structures as shown in Fig. 4.

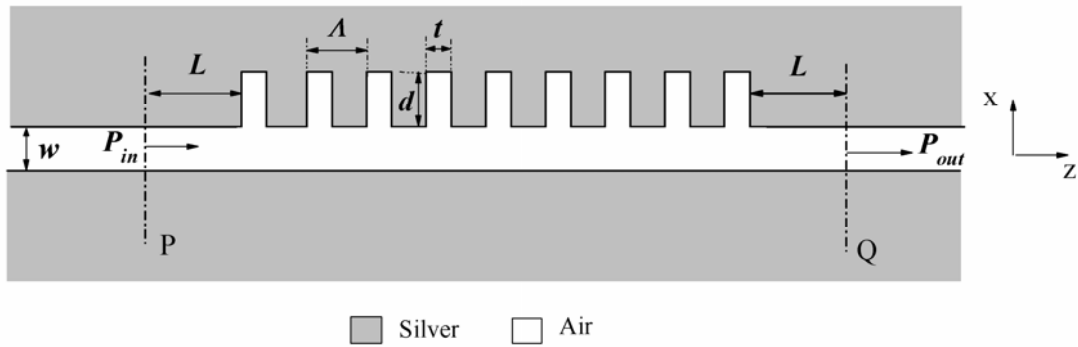

Fig.4. Schematic of a teeth-shaped MDM waveguide filter.

For the multiple teeth-shaped waveguide filter, $w$, $d$ and $t$ stand for the width of the waveguide, height and width of teeth-shaped structure, respectively. $N$ is the number of rectangular teeth and $\Lambda$ is the period of the teeth in the waveguide, the length of $L$ is fixed to be 300nm. According to the scattering matrix theory [29], the SPPs propagation of filter is governed by:

$$\lambda_m = \frac{4 \cdot n_{eff} \cdot d}{(2m+1) - \frac{\Delta\varphi(\lambda)}{\pi}}, (m = 0, 1, 2...), \qquad (4)$$

where $n_{eff}$ is represented as the effective index of the teeth-shaped structure for SPPs, and $\Delta\varphi(\lambda)$ is the phase-shift caused by the reflection. It shows that the central wavelength of the bandgap has a linear relationship with the teeth depth.

We carry on the calculation the transmittance by FDTD for SPPs through the proposed filter by choosing the $w$=50nm, $t$=15nm, $d$=132nm, $N$=9, and a tradeoff value $\Lambda$=215nm to ensure there is no coupling between any two teeth. Fig. 5(a) shows the transmission spectrum which displays the incident plane wave is reflected at the $\lambda$=1.55μm and the corresponding bandgap occurs around this frequency. The transmittance of passband reaches 81% with wavelength range 800nm to 1000nm. When $\Lambda$=100nm is chosen, we can find that the central wavelengths of the bandgaps has a little deviation from $\lambda$=1.55μm due to the coupling of the SPPs waves between two adjacent teeth becomes stronger. The bandgap becomes wider and the passband range 700nm to 1000nm with a high transmittance near 90%. The device dimensions can be 300nm×1000nm is much shorter than the previous MDM grating structures.

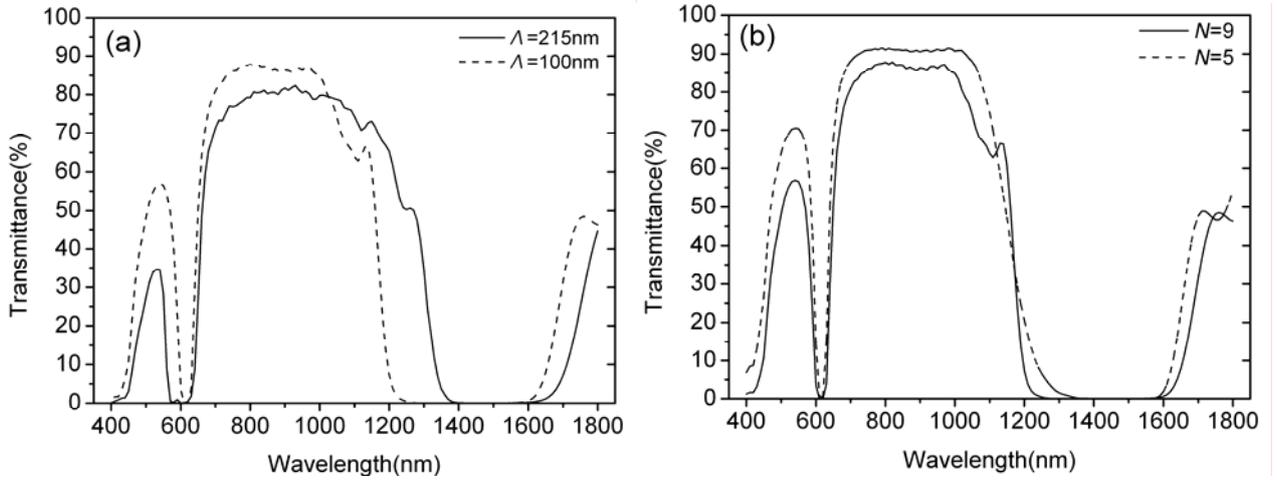

Fig. 5. (a) Transmittance spectra of the teeth filter with $N$=9, (b) Transmittance spectra of the teeth filter consisting of 5 and 9 periods with a fixed $\Lambda$=100nm.

One can see in Fig. 5(b) that, the forbidden bandwidth become narrow with the changing of the period number of $N$ from 9 to 5 with a fixed $\Lambda$=100nm, while the transmittance of the passband increases to 92% with a good performance due to the less propagation loss of the smart structure. Obviously, the period number of 5 is the optimized number with the vertical size less than 1/2 of the free space wavelength to achieve an ultra-compact wavelength filter for high integration.

## 5. CONCLUSIONS

The paper has numerically studied on nano-scale surface plasmon waveguide, Y-branch waveguide and teeth-shaped waveguide filter based on MDM structures. Our results show that the Y-branch waveguide has a large branching angle and a large fabrication tolerance for practical applications. Moreover, a multiple teeth-shaped waveguide proposed not only can realize filtering function of wavelength with a high passband transmittance over 92%, but also has an ultra-compact size in the length of a few hundred nanometers, in comparison with other grating-like SPPs filters. The MDM based Y-branch splitters and teeth-shaped waveguide filters are promise to reduce the difficulties in fabrication, which can be utilized to achieve ultra-compact photonic filtering devices for high integration in SPPs-based flat metallic surfaces.

# ACKNOWLEDGMENT

The authors acknowledge the financial support from the Natural Science Foundation of Guangdong Province, China (Grant No. 07117866).

# REFERENCES

[1] H. Raether, *Surface Plasmon on Smooth and Rough Surfaces and on Gratings* (Springer-Verlag, Berlin, Germany, 1988).
[2] W. L. Barnes, A. Dereux, and T. Ebbesen, "Surface plasmon subwavelength optics," Nature **424**, 824-830 (2003).
[3] E. N. Economou, "Surface plasmons in thin film," Phys. Rev. 182, 539-554 (1969).
[4] L. Liu, Z. Han, and S. He, "Novel surface plasmon waveguide for high integration," Opt. Express **13**, 6645-6650 (2005).
[5] J. A. Dionne, L. A. Sweatlock, and H. A. Atwater, "Plasmon slot waveguides: Towards chip-scale propagation with subwavelength-scale localization," Phys. Rev. B **73**, 035407 (2006).
[6] T. Lee and S. Gray, "Subwavelength light bending by metal slit structures," Opt. Express **13**, 9652-9659 (2005).
[7] G Veronis and S Fan, "Bends and splitters in metal-dielectric-metal subwavelength plasmonic waveguides," Appl. Phys. Lett. **87**, 131102 (2005).
[8] Lin Chen, Bing Wang, and Guo Ping Wang, "High efficiency 90° bending metal heterowaveguides for nanophotonic integration," Appl. Phys. Lett. **89**, 243120 (2006).
[9] H. Gao, H. Shi, C. Wang, C. Du, X. Luo, Q. Deng, Y. Lv, X. Lin, and H. Yao, "Surface plasmon polariton propagation and combination in Y-shaped metallic channels," Opt. Express **13**, 10795-10800 (2005).
[10] Z. Han and S. He, "Multimode interference effect in plasmonic subwavelength waveguides and an ultra-compact power splitter," Opt. Commun. **278**, 199-203 (2007).
[11] T. Nikolajsen, K. Leosson and S. I. Bozhevolnyia, "Surface plasmon polariton based modulators and switches operating at telecom wavelengths," Appl. Phys. Lett. **85**, 5833 (2004).
[12] H. Zhao, X. Huang and J. Huang, "Novel optical directional coupler based on surface plasmon polaritons," Phys. E **40**, 3025-3029 (2008).
[13] B. Wang and G. Wang, "Surface plasmon polariton propagation in nanoscale metal gap waveguides," Opt. Lett. **29**, 1992-1994 (2004).
[14] Z. Han, L. Liu, and E. Forsberg, "Ultra-compact directional couplers and Mach-Zehnder interferometers employing surface plasmon polaritons," Opt. Commun. **259**, 690-695 (2006).
[15] H. Ditlbacher, J. R. Krenn, G. Schider, A. Leitner, and F. R. Aussenegg, "Two-dimensional optics with surface plasmon polaritons," Appl. Phys. Lett. **81**, 1762 (2002).
[16] S. I. Bozhevolnyi, J. E. Erland, K. Leosson, P. M. W. Skovgaard, and J. M. Hvam, "Waveguiding in surface plasmon polariton band gap structures," Phys. Rev. Lett. **86**, 3008-3011 (2001).
[17] Z. Han, E. Forsberg, and S. He, "Surface plasmon Bragg gratings formed in metal-insulator-metal waveguides," IEEE Photon.Technol. Lett. **19**, 91-93 (2007).
[18] T. Holmgaard, Z. Chen, S. I. Bozhevolnyi, L. Markey, A. Dereux, A. V. Krasavin, and A. V. Zayats, "Bend and splitting loss of dielectric-loaded surface plasmon-polariton waveguides," Opt. Express **16**, 13585-13592 (2008).
[19] A. V. Krasavin and A. V. Zayats, "Passive photonic elements based on dielectric-loaded surface plasmon polariton waveguides," Appl. Phys. Lett. **90**, 211101 (2007).
[20] S. Passinger, A. Seidel, C. Ohrt, C. Reinhardt, A. Stepanov, R. Kiyan, and B. Chichkov, "Novel efficient design of Y-splitter for surface plasmon polariton applications," Opt. Express **16**, 14369-14379 (2008).
[21] E. D. Palik, *Handbook of optical constants of solids* (Academic Press, New York, NY 1985).
[22] B. Wang and G. P. Wang, "Plasmon Bragg reflector and nanocavities on flat metallic surfaces," App. Phys. Lett. **87**, 013107 (2005).
[23] W. Lin and G. Wang, "Metal heterowaveguide superlattices for a plasmonic analog to electronic Bloch oscillations," Appl. Phys. Lett. **91**, 143121 (2007).
[24] L. Zhou, X. Yu, and Y. Zhu, "Propagation and dual-localization of surface plasmon polaritons in a quasiperiodicmetal heterowaveguide," App. Phys. Lett. **89**, 051901 (2006).
[25] A. Hossieni and Y. Massoud, "A low-loss metal-insulator-metal plasmonic bragg reflector," Opt. Express **14**, 11318-11323 (2006).

[26] A. Hosseini, H. Nejati, and Y. Massoud, "Modeling and design methodology for metal-insulator-metal plasmonic Bragg reflectors," Opt. Express. **16**, 1475–1480 (2008).
[27] J. Park, H. Kim, and B. Lee, "High order plasmonic Bragg reflection in the metal-insulator-metal waveguide Bragg grating," Opt. Express **16**, 413-425 (2008).
[28] J. Q. Liu, L. L. Wang, M. D. He, W. Q. Huang, D. Y Wang, B. S. Zou, and S.C Wen, "A wide bandgap plasmonic Bragg reflector," Opt. Express. **16**, 4888–4894 (2008).
[29] H. A. Haus, *Waves and Fields in Optoelectronics* (Prentice-Hall, Englewood Cliffs, NJ, 1984).